\documentstyle[12pt,epsf]{article}
\begin {document}

\begin{center}

{\bf Non-thermalizability of a Quantum Field Theory}\\

\bigskip

C. R. Hagen*\\

\bigskip

Department of Physics and Astronomy\\
University of Rochester\\
Rochester, N.Y. 14627\\

\end{center}
\begin{abstract}

     The problem of understanding the role of large gauge transformations in
thermal field theories has recently inspired a number of studies of a one
dimensional field theory.  Such work has led to the conclusion that gauge
invariance is restored only when the entire perturbation expansion can be
summed.  A careful reexamination of that model is shown, however, to lead to
vastly different conclusions when the constraint implied by the field equations
is explicitly taken into account.  In particular it is found that none of the
relevant propagators has any temperature dependence and that the effective
action is essentially trivial.  A generalization of the model to include bosons
as well as fermions is also solved with qualitatively identical results being
obtained.  

\end{abstract}

\bigskip

     A cause of considerable concern in the analysis of Chern-Simons field
theories in which a coupling constant quantization mechanism is at work is the
fact that induced correction terms are found to be temperature dependent.  In
other words when such a field theory is thermalized the effective Chern-Simons
term is found to have a coefficient which varies smoothly with the temperature,
most typically through a term of the form $\tanh({\beta M/2})$.  Since the
argument for coupling constant quantization is essentially based upon large
gauge transformations, such temperature dependence suggests that the
calculational program does not preserve gauge invariance under this class of
transformations.

     This problem has recently been addressed in a much cited paper [1] which
considers a (0 + 1) dimensional field theory of $N_f$ fermions $\psi_j, 
j=1,...,N_f$ minimally coupled to a $U(1)$ gauge field $A$ [2].  In terms of a
real time formulation the Lagrange density is 

\begin{equation}
{\cal L} = \sum^{N_f}_{j=1}\psi^\dagger_j(i\partial_t - m - A)\psi_j +
\kappa A
\end{equation}

\noindent with $m$ being referred to as a mass parameter.  The last term in
(1) is a linear Chern-Simons term appropriate to a (0 + 1) dimensional theory. 
The Lagrangian implies the equation of motion 

\begin{equation}
(i\partial_t - m - A)\psi_j=0
\end{equation}

\noindent and the constraint equation

\begin{equation}
\kappa = Q
\end{equation}

\noindent where[3]

$$Q=\sum^{N_f}_{j=1}\psi^\dagger_j\psi_j.$$

The only nonvanishing equal time (anti)commutation relation is given by 

\begin{equation}
\{ \psi_i,\psi^\dagger_j \}=\delta _{ij}
\end{equation}
 
\noindent with $\psi_j$ having the additional property that it annihilates the
ground state $|0>$, i.e., 

\begin{equation}
\psi_j|0>=0.
\end{equation}

\noindent Although Eq.(3) clearly cannot be valid in the entire Fock space
associated with the operators $\psi_j$, it can be realized on a reduced physical
state space of fixed charge $\kappa$.  This leads to the further conclusion
that $\kappa$ can assume only positive integer values $n $ as a consequence of
the constraint (3) together with (4) and (5).  Specifically, for an arbitrary
state $|>$ Eqs.(3) and (4) imply that the equation 

$$\kappa |>=Q|>$$
is consistent only for states $|>$ which have integer $n$ total charge with $n$
equal to the allowed value of $\kappa$[4].  Thus the $\kappa$ quantization of
this work and that of ref. 1 have totally different origins; namely, it follows
from the constraint equation here and from topological arguments in the latter.
  
  It was found in ref. 1 that the thermalized version of this theory yields an
effective action which is a highly nontrivial function of the temperature, the
$m$ parameter, and the integral over $A$.  It has the property that its 
expansion in powers of $A$ has the continuous dependence upon temperature
typical of (2+1) dimensional perturbation theory even though the exact version
is consistent with gauge invariance.  This result has been the basis of a 
number of studies in (2+1) dimensions [5-8] which depend on specially
constructed models which more or less factorize into a (0+1)
dimensional part and a two dimensional Euclidean one.  Despite the fact that
all those results seem to be eminently reasonable, it will be seen here that
the (0+1) dimensional model upon which they all ultimately depend is found to
have a totally different solution when the constraint (3) is taken into
account.   

     To demonstrate the claimed result it is well to point out at the outset
that the model (1) is properly described as a Schr\"odinger model (or better, a
Galilean invariant one) rather than as a (0+1) dimensional theory in the
context of special relativity.  This conclusion is hard to avoid in view of
the fact that there are no antiparticles in the theory and consequently no
charge conjugation operator.  Once this observation is made one is naturally
led to the interpretation of $m$ not as a mass parameter, but rather as an
internal energy.  This means that there is an energy \lq\lq barrier" $m$ which
has to be overcome in the creation of each fermion.  Since, however, only
states of charge $n$ are allowed by the constraint, the total energy barrier
$nm$ amounts merely to a trivial redefinition of the zero point of energy[9].
In terms of operators one can make the change of variable 

$$\psi'_j(t)=e^{imt}\psi_j(t)$$
thereby totally eliminating dependence upon the parameter $m$.  Since the charge
operator is also independent of $m$, one is left with the apparently
paradoxical situation that the effective action is required by
general considerations to be $m$ independent even though explicit calculations
based upon thermal field theory[1,2] lead to the opposite conclusion.
It remains to be shown how one can eliminate this seeming contradiction.   

     One begins the calculation of the effective action with an evaluation of
the various fermionic propagators.  These are defined in terms of the time
ordered products by 

$$S(t,t^\prime)=i\epsilon (t,t^\prime)
{Tr e^{-iHT}(\psi_j(t)\psi^\dagger_j(t^\prime))_+\over
Tre^{-iHT}}   $$
where $0\leq t,t'\leq T$ and $\epsilon (t,t')$ is the alternating function. 
The trace is to be taken over the space of physical states, namely those of
fixed total charge $n$.  By letting $T\to-i\beta$ where $\beta$ is the
reciprocal temperature one obtains the usual thermal propagator.  Note also
that the index $j$ on $S$ has been suppressed since the fields $\psi_j$ all have
the same $m$ value.  Upon using the form

$$H=m\sum^{N_f}_{j=1}\psi^{\dagger}_j\psi_j$$
for the Hamiltonian and the fact that for all allowed states of the system
$H$ has the constant value $nm$ the expression for the propagator reduces to
the $T$ independent form

$$S(t,t')=i\epsilon 
(t,t'){Tr (\psi_j(t)\psi^{\dagger}_j(t'))_+\over Tr 1}.$$
Physically, this $T$ independence arises, of course, from the fact that the
thermal averaging is carried out only over a set of states all of which are
degenerate in energy with each other.  Thus the constraint leads directly to
the claimed result that this theory cannot be thermalized in the usual fashion.
It may be noted that if one allows the various particles in the theory to have
different $m$ (i.e., internal energy) values, a $T$ dependence can be forced
into the model.  However, the effective action will not display such $T$
dependence and will in fact coincide with the degenerate $m$ case.  

     It is clearly of interest to explain at this point why the usual
thermalized propagators are not valid in the present case.  To this end it
should be noted that the standard derivation depends upon two assumptions,
neither of which is valid for this model.  The first of these is that the
Hamiltonian generates the time dependence of the fermion operators; namely,
that 

$$\psi_j(t)=e^{iHt}\psi_j(0)e^{-iHt}.$$
However, this equation is certainly incorrect in the application at hand since
it fails to reproduce the correct equation of motion.  More specifically, one
notes that $\psi_j$ depends on the integral of $A(t)$ and thus the commutator of
$H$ with $\psi_j$ fails to give its time derivative.  
     A second reason for the failure of the usual derivation of the thermal
propagator is that it requires the validity of the expression 

\begin{equation}
Tr AB=Tr BA
\end{equation}
which is not tenable for this model.  The reason is that although bilinear
operators such as the charge are well defined, one violates (6) in giving
meaning to the operator $\psi_j$ which has the property of taking one out of
the space of physical states.  Thus one finds that any attempt to derive the
usual thermal boundary conditions by application of Eq.(6) in effect transforms
the trace over the physical state space into one over an unphysical set of
states.

     To calculate the propagator one begins with the $A=0$ case which will be
designated by $S_0(t,t')$.  It can be written as
                                               
$$S_0(t,t')=i\theta_+(t,t')<0|\psi_j(t)\psi^{\dagger}_j(t')|0>-i\theta_-(t,t')
<0|\psi^\dagger _j(t')\psi_j(t)|0>$$
where $\theta _\pm(x)={1\over2}(1\pm {x\over |x|})$.  This readily leads to the
result

\begin{equation}
S_0(t,t')=i[\theta_+(t,t')-{n\over N_f}]e^{-im(t-t')}.
\end{equation}
It may be noted that this differs from ref. 2 only in that in the
latter case $n$ is a function of temperature while here it is simply a
nonnegative integer.

     The effect of the interaction can now be included by considering the
equation for the $A\neq 0$ propagator

$$(-i\partial_t +m)S(t,t')=\delta(t-t')-A(t)S(t,t').$$
Although it might be supposed that this equation is to be solved perturbatively
by expansion of the formal expression

\begin{equation}
S=S_0(1+AS_0)^{-1}
\end{equation}
with $S_0$ given by Eq.(7), this is yet another instance in which the usual
assumptions fail.  The essential point here is that correct boundary conditions
on the exact propagator cannot be ensured by the imposition of an appropriate
set of boundary conditions on the inverse of the operator $-i\partial_t +m$,
thereby leading to an acceptable solution of the form (8).  This stands in
marked contrast to the case of relativistic field theory in which the
invocation of causal boundary conditions (i.e., positive frequencies in the
future, negative frequencies in the past) on the non-interacting propagators
automatically ensures that such conditions will characterize the entire
perturbation series.  Similarly, the imposition of causal (i.e., retarded) or
anti-causal (advanced) boundary conditions in a Galilean theory will suffice to
guarantee that they will be satisfied by the exact solution.  

     One obtains a solution for the exact propagator in either of two different
ways.  The more basic approach merely recognizes that the equation for this
function is simply a first order differential equation in a single variable 
which must therefore be soluble by elementary means.  The result is

\begin{equation}
S(t,t')=S_0 (t,t')exp{\left[-i\int^t_{t'} A(\tau )d\tau \right]}.
\end{equation}
Alternatively, one could derive (9) by an approach which is close in spirit to
the result (8).  However, it is first necessary to split $S(t,t')$ 
into two parts
so that proper boundary conditions can be imposed.  Thus one defines 

$$S_\pm (t,t')=\theta_\pm (t,t')S(t,t')$$
and notes that these functions satisfy the equations

$$(-i\partial_t +m+ A)S_+(t,t')=(1-{n\over {N_f}})\delta(t-t')$$
and
$$(-i\partial_t +m+ A)S_-(t,t')= {n\over N_f} \delta (t-t').$$
Since $S_+(t,t')$ and $S_-(t,t')$ clearly must satisfy retarded and advanced
boundary conditions respectively, it is now straightforward to obtain solutions
of these equations as 

\begin{equation}
S_+=(1-{n\over N_f})S_{0+}(1+AS_{0+})^{-1}
\end{equation}
and

\begin{equation}
S_-={n\over N_f}S_{0-}(1+AS_{0-})^{-1}
\end{equation}
where

$$S_{0\pm} (t,t')=\pm i\theta_{\pm}(t,t')e^{-im (t-t')}.$$

     Upon combining the solutions $S_\pm$ one obtains the result

$$S(t,t')=S_+(t,t') + S_-(t,t').$$ 
From Eq.(10) it follows that

\begin{eqnarray*}
S_+(t,t') & = & (1-{n\over N_f}) S_{0+}(t,t')\sum^\infty _{r=0}(-i)^r
\int d\tau_1\int d\tau_2...
\int d\tau_r \; A(\tau_1)A(\tau_2)...A(\tau_r)  \\
&  & \theta_+(t-\tau_1)\theta_+(\tau_1-\tau_2)...\theta_+(\tau_r-t').
\end{eqnarray*}
However, this is clearly equivalent to 

$$S_+(t,t')=(1-{n\over N_f})S_{0+}(t,t')
exp\left[-i\int^t _{t'}A(\tau)d\tau \right].$$
Corresponding manipulation of Eq.(11) yields

$$S_-(t,t')={n\over N_f}S_{0-}(t,t')
exp\left[-i\int^t _{t'}A(\tau)d\tau\right].$$
Upon combining these results for $S_{\pm}(t,t')$ one obtains the
result (9), thereby demonstrating the equivalence of the two approaches.

     It is now straightforward to obtain the effective action as a function of
$A$.  To this end one notes that 

$$<Q>=iN_f\lim _{\epsilon\to 0}S(t,t'=t+\epsilon )$$ 
which from (7) and (9) is readily seen to yield

$$<Q>=n$$
in agreement with the constraint (3).  Thus the effective action $\Gamma (A)$
has the form 

$$\Gamma (A)=in\int A(\tau )d\tau,$$
a result which stands in marked contrast to what has been obtained in 
treatments which do not take into account the constraint (3).

     The model considered here can be extended to include bosons as well as
fermions[10].  Such a system is obtained by the replacement of (1) by

\begin{equation}
{\cal L}=\sum^{N_f}_{j=1}\psi^{\dagger}_j(i\partial_t-m-A)\psi_j +
\sum^{N_b}_{j=1}\phi^{\dagger}_j(i\partial_t-\mu -A)\phi_j +\kappa A.
\end{equation}
The boson operators $\phi_j, j=1,...,N_b$ have the nonvanishing equal time
commutators
$$[\phi_i,\phi^{\dagger}_j]=\delta_{ij}$$
and the property that 
$$\phi_j|0>=0.$$
The total charge operator $Q$ now has the form

\begin{equation}
Q=\sum^{N_f}_{j=1}\psi^{\dagger}_j\psi_j +
\sum^{N_b}_{j=1}\phi^{\dagger}_j\phi_j
\end{equation}
with the constraint (3) being unmodified when expressed in terms of the charge
(13).  As before one concludes that $\kappa$ must be a nonnegative integer $n$,
although unlike the purely fermionic case it is no longer bounded from above. 

     By assuming that the fermionic and bosonic internal energies $m$ and $\mu$
are unequal one can obtain propagators which have a temperature dependence[11].
Despite such dependence the arguments against using the usual thermal
propagators remain valid and one obtains the correct result by combinatorial
arguments.  Thus for the $A=0$ fermionic propagator $S_0(t,t')$ one finds
that the Hamiltonian

$$H=m\sum^{N_f}_{j=1}\psi^{\dagger}_j\psi_j+\mu
\sum^{N_b}_{j=1}\phi^{\dagger}_j\phi_j$$
implies the result 

$$S_0(t,t')=i\left[\theta_+(t,t')-{i\over N_f}
{\partial\over{\partial(mT)}}log C\right] e^{-im(t-t')}$$
where 

$$C=\sum^N_{r=0}e^{-i[rm+(n-r)\mu]T}{N_f!\over r!(N_f-r)!}
{(N_b+n-r-1)!\over (N_b-1)!(n-r)!}$$
where $N$ is the lesser of $n$ and $N_f.$   
     The $\phi$ propagators $D(t,t')$ are defined as

$$D(t,t')=i{Tre^{-iHT}(\phi_j(t)\phi^{\dagger}_j(t'))_+\over Tre^{-iHT}}$$
and are found in the $A=0$ limit to be 

$$D_0(t,t')=i\left[\theta(t,t')+ {1\over N_b}
\left(n-i{\partial \over \partial(mT)}log C\right)\right]
e^{-i\mu (t-t')}.$$ 
As before one finds that 

$$S(t,t')=S_0(t,t')exp\left[-i\int^t_{t'} A(\tau) d\tau \right]$$
and

$$D(t,t')=D_0(t,t')exp\left[-i\int^t_{t'} A(\tau) d\tau \right].$$
This implies that the matrix element of the charge operator as given by

$$<Q>=iN_f\lim_{\epsilon \to 0}S(t,t'=t+\epsilon )-iN_b\lim_{\epsilon
\to 0}D(t,t'=t+\epsilon)$$
reduces to

$$<Q>=n$$
so that the effective action is the same as in the purely fermionic case.
     Thus one finds that the inclusion of bosonic fields, while providing a
natural mechanism for the introduction of temperature dependence into the
various propagators, has the property of leaving the matrix elements of the
charge operator (and hence the effective action) unchanged and temperature
independent just as in the original model.  One is forced to conclude that
the laudable goal of finding a mechanism to explain the disturbing temperature
dependence in (2+1) dimensional Chern-Simons theories is not to be realized in
the thermalization of the model (1).  

\bigskip

The author acknowledges useful discussions with Dr. O. Kong.
This work is supported in part by the U.S. Department of Energy Grant
No.DE-FG02-91ER40685.

\medskip

*Electronic address: hagen@urhep.pas.rochester.edu

\medskip

\newpage

\noindent References

\begin{enumerate}
\item G. Dunne, K. Lee, and C. Lu, Phys. Rev. Lett. {\bf 78}, 3434 (1997).
\item Additional study of this model is to be found in A. Das and G. Dunne,
hep-th/9712144.

\item It should be noted that the definition of $Q$ in ref. 1 differs from that
used here by the term $N_f/2$ so that the charge of the state $|0>$ has the more
conventional value of zero in the present work.  

\item Because of the fermionic statistics $n$ must satisfy the condition 
$0\leq n \leq N_f$.

\item S. Deser, L. Griguolo, and D. Seminara, Phys. Rev. Lett. {\bf 79},
1976 (1997).

\item C. D. Fosco, G. L. Rossini, and F. A. Schaposnik, Phys. Rev. Lett. {\bf
78}, 1980 (1997).

\item I. J. R. Aitchison and C. Fosco, hep-th/9709035.

\item R. Gonz\'alez-Felipe, hep-th/9709079.

\item In the event that particles of different masses can transmute into each
other, the differences between the zero points of internal energy for various
particles can have observable significance.  An example of this occurs in the
Galilean Lee model which has been considered by J. M. L\'evy-Leblond, Commun.
math Phys. {\bf 6}, 286 (1967).

\item  The study of such a system was in fact suggested in ref.1.

\item Just as in the fermionic case all the internal energies can be
transformed away.  This will not be done here in order to emphasize the fact
that the effective action is temperature independent regardless of whether one
carries out such a step.
\end{enumerate}

\end{document}